\documentclass[final,3p,times,twocolumn]{elsarticle}

\usepackage{amssymb}
\usepackage{array}
\usepackage{amsmath}
\usepackage{makecell}
\usepackage[utf8]{inputenc}
\usepackage{subfigure}
\usepackage{float}
\usepackage{xcolor}
\PassOptionsToPackage{hyphens}{url}\usepackage{hyperref}

\newcommand{\paper}{paper}
\newcommand{\target}{passive}

\usepackage{amsthm}

\journal{Robotic and Autonomous Systems}

\begin{document}
	
	\begin{frontmatter}

		\author{Fabrizia~Auletta\fnref{label1,label3}}
		\author{Davide~Fiore\fnref{label2}}
		\author{Michael~J.~Richardson\fnref{label3}}
		\author{Mario~di~Bernardo\corref{cor1}\fnref{label4}}
	
		\fntext[label1]{Fabrizia Auletta is with the Department of Engineering Mathematics, University of Bristol, University Walk, BS8 1TR Bristol, U.K. and also with the Department of Psychology, Faculty of Medicine, Health and Human Sciences, Macquarie University, Sydney, NSW 2109, Australia. {\tt\small fabrizia.auletta@bristol.ac.uk}}
		\fntext[label2]{Davide Fiore is with the Department of Mathematics and Applications ``R. Caccioppoli", University of Naples Federico II, Via Cintia, Monte S. Angelo, 80126 Naples, Italy. {\tt\small  davide.fiore@unina.it}}
		\fntext[label3]{Michael J. Richardson is with the Department of Psychology, Faculty of Medicine, Health and Human Sciences, and the Center for Elite Performance, Expertise and Training, Macquarie University, Sydney, NSW 2109, Australia. {\tt\small michael.j.richardson@mq.edu.au}}
		\fntext[label4]{Mario di Bernardo is with the Department of Electrical Engineering and Information Technology, University of Naples Federico II, Via Claudio 21, 80125 Naples, Italy. {\tt\small mario.dibernardo@unina.it}}
		\cortext[cor1]{\textit{Corresponding author: Mario di Bernardo}}     
		
		\title{Herding stochastic autonomous agents via local control rules and online global target selection strategies}

		\begin{abstract}
			In this \paper{} we propose a simple yet effective set of local control rules to make a group of ``herder agents'' collect and contain in a desired region an ensemble of non-cooperative stochastic ``target agents'' in the plane. 
			We investigate the robustness of the proposed strategies to variations of the number of target agents and the strength of the repulsive force they feel when in proximity of the herders. 
			Extensive numerical simulations confirm the effectiveness of the approach and are complemented by a more realistic validation on commercially available robotic agents via ROS.
		\end{abstract}
		
		%%Graphical abstract
		%\begin{graphicalabstract}
		%\includegraphics{grabs}
		%\end{graphicalabstract}
		
		%%%Research highlights
		%\begin{highlights}
		%	\item Research highlight 1
		%	\item Research highlight 2
		%\end{highlights}

		
	\end{frontmatter}

\section{Introduction}
\label{sec:intro}
Exploration and rescue, evacuation from dangers, surveillance and crowd control are all examples of \emph{multi-agent herding problems} in which two kinds of agents interact \cite{murphy_Rescue, trautman_crowd}.
In these problems, a set of ``active'' agents (the herders) need to drive a set of ``passive'' agents (the herd) towards a desired goal region and confine them therein. 
In most cases, repulsive forces exerted by the herders on the herd are exploited to drive the movements of the \target{} agents that need to be corralled and, at times, cooperation among the herders (such as attractive forces between them) are used to enhance the herding performance. 
Notable herding solutions are those proposed in \cite{vaughan_experiments_2000, jyh-ming_lien_shepherding_2004, strombom_solving_2014, Paranjape_2018, licitra2019single} for single herders and in \cite{jyh-ming_lien_shepherding_2005, Hacque2009, wonki_autonomous_2017, pierson_controlling_2018, nalepka_first_2017} for multiple herders.

One of the problems to be addressed in the control design of herder agents is deciding at any given time what \target{} agent a herder should target first when more than one herder is present.
For the sake of comparison with our approach, we now briefly review the most relevant research from the literature addressing multi-agent herding, where more than one herder is required to collect and drive a group of \target{} agents towards a desired goal region. 
\paragraph{Related work}
\label{sec:RelWorks}
One of the earliest solutions to the herding problem was proposed by Lien \textit{et al.} in \cite{jyh-ming_lien_shepherding_2004} and \cite{jyh-ming_lien_shepherding_2005}. The trajectories followed by \target{} and herder agents were generated using global rule-based roadmaps -- abstract representations of the walkable paths given as a directed graph \cite{Wilmarth1999}. 
Numerical simulations showed that multiple herders were successful in coping with increasing sizes of the herd. Nevertheless, herders' performance worsened as the flocking tendency of \target{} agents decreased. 
	
Multi-agent herding scenarios were also considered in \cite{Hacque2009,Haque2011}. Here the authors addressed the problem of controlling a group of herders so as to entrap a group of \target{} agents in a region from which they could not escape. To solve this problem, each herder was pre-assigned some region of influence. Targets' motion was then only influenced by a specific herder if they happened to be within its region of influence; travelling otherwise at constant speed and heading aligned to that of their neighbouring agents. The velocities of the herders were regulated according to that of the other \target{} agents with which they interacted, arranging themselves in two opposite rows or in a carousel.
	
Other multi-agent herding scenarios where many herders are required to collect and patrol a group of \target{} agents were also proposed in \cite{wonki_autonomous_2017}. Inspired by the limited visual field of real sheepdogs and the absence of centralised coordination among them, the latter work proposed a herding algorithm based entirely on local control rules. The dynamics of both herders and \target{} agents were modelled as the linear combination of potential field-like forces within a sensing area.
In addition to this basic dynamics, \target{} agents were also subject to a repulsive force from the herders. Herders were controlled by an appropriate input selected as a function of their distance from the nearest \target{} agent and their distance from a desired goal. 
The result of the proposed shepherding behaviour was the emergence of an arc formation among the herders (a similar formation was instead hard-coded in the algorithm presented earlier in \cite{jyh-ming_lien_shepherding_2005}). Numerical simulations showed the effectiveness of the approach under the assumption that \target{} agents tend to flock together. In this case, herders could indeed collect and herd multiple sub-flocks without any explicit coordination rule. 

%Pierson
In Robotics, feedback control strategies have been recently presented to solve multi-agent herding problems and guarantee convergence of the overall system.
For instance, in \cite{pierson_controlling_2018} the case of multiple herder agents regulating the mean position of a group of flocking \target{} agents was investigated. An arc-based strategy was proposed for the herders to surround and drive the targets towards a desired goal region. The proposed control law and its convergence properties were explored by modeling the whole herd as a single unicycle controlled by means of a point-offset technique (see \cite{pierson_controlling_2018} for further details).

A different approach was used in Cognitive Science \cite{nalepka_investigating_2015, nalepka_first_2017, nalepka_herd_2017, Nalepka_2019}, where a model of the herding agent was derived from experimental observations of how two human players herd a group of randomly moving agents in a virtual reality setting. 
It was observed that, at the beginning of the task, all pairs of human players adopted a \emph{search and recovery} strategy; players individually chasing the farthest \target{} agent in the half of the game field assigned to them and driving it inside the desired containment region.
Once all agents are gathered inside the goal region, most pairs of human herders were observed to switch to an entirely different containment strategy, based on exhibiting an oscillatory movement along an arc around the goal region creating effectively a ``repulsive wall'' for the \target{} agents keeping them therein \cite{nalepka_herd_2017}. 
To reproduce this behaviour in artificial agents, a nonlinear model was proposed in \cite{Nalepka_2019} where the switch from search and recovery to the oscillatory containment strategy is induced by a Hopf bifurcation triggered by a change in the distance of the herd agents from the goal region.

With regard to a \emph{single} herder agent gathering one-by-one a group of \target{} agents, recent work by \cite{licitra_single_2017} employed a backstepping control strategy for the single herder to chase one target at a time, with the herder switching among different targets and succeeding in collecting them within a goal region of interest. This idea was further developed in \cite{licitra_single_2018,licitra2019single} where other control strategies and further uncertainties in the herd's dynamics were investigated.
An alternative approach is to frame the problem as a pursuit-evasion game, as done for example in \cite{kachroo_dynamic_2001,zuazua2016,deptula_single_2018}, where the case of one \target{} agent evading from one pursuer is solved by computing off-line the optimal solution of a dynamic programming problem; the case of multi-driver and multi-evader agents being more recently analysed in \cite{ko2019asymptotic}.

\subsection{Contributions of this paper}
In this \paper{}, we consider the case of multiple herders chasing a group of \target{} agents whose dynamics, as often happens with natural agents such as fish, birds or bacteria, is stochastic and driven by a random Brownian noise. 
However, contrary to what is usually done in the rest of the literature \cite{Hacque2009, jyh-ming_lien_shepherding_2004, pierson_controlling_2018,wonki_autonomous_2017}, we do not consider the presence of any flocking behaviour between \target{} agents, making the problem more complicated to solve as each target needs to be tracked and collected independently from the others. 

To solve the problem, we present a simple, yet effective, dynamic herding strategy based on the combination of local feedback control laws among the agents and a set of global target selection rules that drive how herders make decisions on what targets to follow.
With respect to other solutions in the literature \cite{jyh-ming_lien_shepherding_2004,pierson_controlling_2018}, our approach does not involve the use of \emph{ad hoc} formation control strategies to force the herders surround the herd, but we rather enforce cooperation between herders by dynamically dividing the plane among them by means of simple yet effective and robust rules that can be easily implemented in real robots.

We then numerically analyse how robust these strategies are to parameter perturbations, uncertainties and unmodeled disturbances in \target{} agent dynamics.
Moreover, we assess how different choices of the target selection rules affect the overall effectiveness of the methodology we propose. 
Finally, for the sake of completeness we provide a ROS implementation of our strategy to test its ability to solve the herding problem in a more realistic robotic setting.

\section{The herding problem}
\label{sec:ProbFormulation}
\begin{figure}[!t]
\centering
\includegraphics[width=0.8\linewidth]{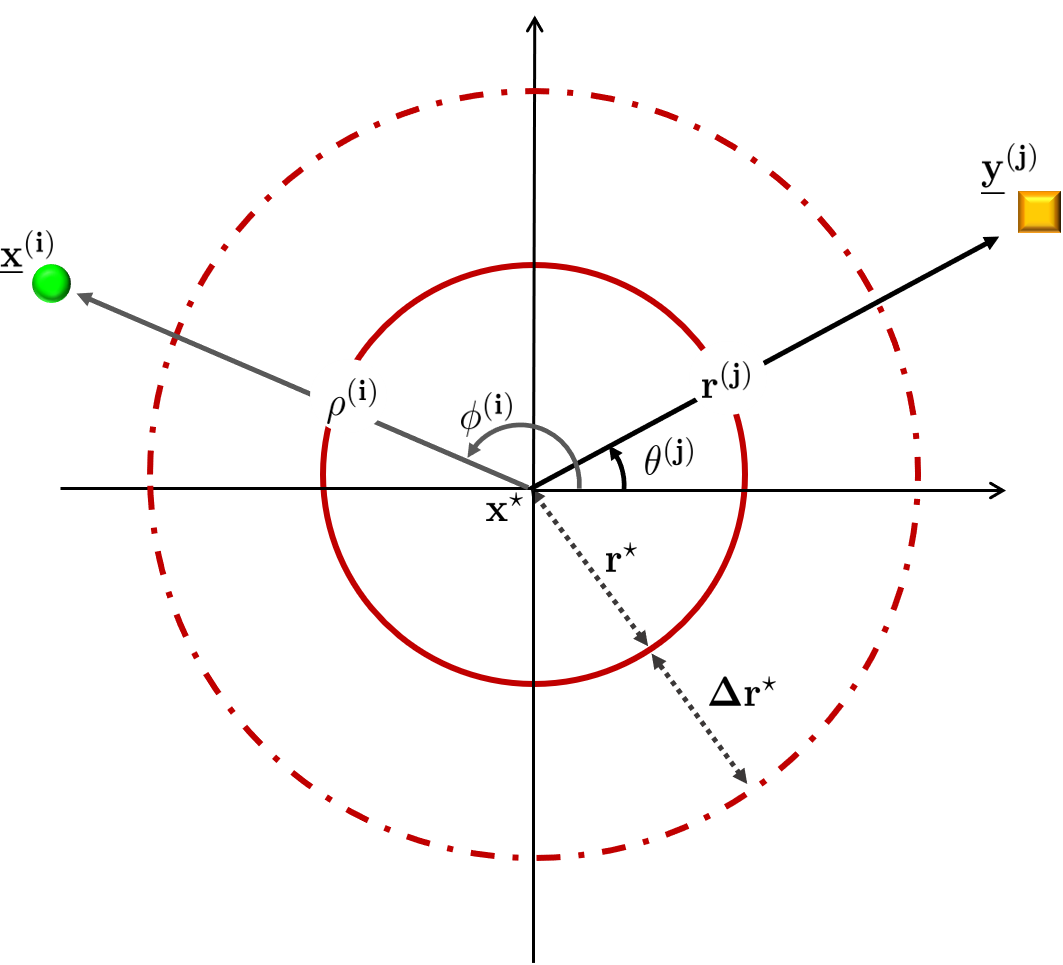}
\caption{Illustration of the spatial arrangement in the herding problem. The herder agent $ \underline y^{(j)} $ (yellow square), with polar coordinates $( r^{(j)},\,\theta^{(j)} )$, must relocate the target agent $\underline x ^{(i)}$ (green ball), with polar coordinates $(\rho^{(i)},\, \phi^{(i)}) $, in the containment region $\mathcal{G}$ (solid red circle) of centre $ \underline{x}^{\star} $ and radius $ r^{\star} $. The buffer region $\mathcal{B}$, of width $\Delta r^\star$, is depicted as a dashed red circle.}
\label{fig:geometry}
\end{figure}

We consider the problem of controlling ${N}_H\geq 2$ herder agents in order for them to drive a group of $N_T > N_H$ \target{} agents in the plane ($\mathbb{R}^2$) towards a goal region and contain them therein. 
We term $\underline{y}^{(j)}$ the position in Cartesian coordinates of the $j$-th herder in the plane and $\underline{x}^{(i)}$ that of the $i$-th \target{} agent. 
We denote as $(r^{(j)},\, \theta^{(j)})$ and $(\rho^{(i)},\, \phi^{(i)})$ their respective positions in polar coordinates as shown in Fig.~\ref{fig:geometry}. 
We assume the goal of the herders is to drive the \target{} agents towards a circular \emph{containment region} $\mathcal{G}$, of radius $r^\star$ centred at $\underline{x}^{\star}$.
Without loss of generality, we set $\underline{x}^{\star}$ to be the origin of $\mathbb{R}^2$. 

Assuming the herders have their own trivial dynamics in the plane, the \emph{herding problem} can be formulated as the design of the control action $u$ governing the dynamics of the herders given by
\begin{equation}
\label{eq:herder} 
m\, \underline{\ddot{y}}^{(j)}=u(t,\underline{x}^{(1)}, \dots, \underline{x}^{(N_T)},\underline{y}^{(1)},\dots, \underline{y}^{(N_H)}),
\end{equation}
where $m$ denotes the mass of the herders assumed to be unitary, so that the herders can influence the dynamics of the \target{} agents (whose dynamics will be specified in the next section) and guarantee that
\begin{equation*}
\label{eq:goal}
\|\underline{x}^{(i)}(t) - \underline{x}^{\star}\| \leq r^{\star}, \qquad \forall i, \forall t\geq \bar{t},  
\end{equation*}
where $\lVert \cdot \rVert$ denotes the Euclidean norm; that is, all \target{} agents are contained, after some finite time $\bar{t}$, in the desired region $\mathcal{G}$.

We assume an annular \emph{safety region} $\mathcal{B}$ of width $\Delta r^\star$ exists surrounding the goal region that the herders leave between themselves and the region where targets are contained.

In what follows we also assume that (i) herder and \target{} agents can move freely in $\mathbb{R}^2$; (ii)
herder agents have global knowledge of the environment and of the positions of the other agents therein.

\section{Target dynamics}
\label{sec:TargetModel}
Taking inspiration from \cite{nalepka_first_2017}, we assume that, when interacting with the herders, \target{} agents are repelled from them and move away in the opposite direction, while in the absence of any external interaction, they randomly diffuse in the plane.
Specifically, we assume \target{} agents move according to the following stochastic dynamics
\begin{equation}
\label{eq:target_nalepka}
d \underline x^{(i)}(t) =  V^{(i)}_r(t) dt + \alpha_b d W^{(i)}(t), 
\end{equation}
where $V^{(i)}_r(t)$ describes the repulsion exerted by all the herders on the $i$-th \target{} agent, 
$W^{(i)}(t)=[W^{(i)}_1(t), W^{(i)}_2(t)]^\top$ is a 2-dimensional standard Wiener process and $\alpha_b>0$ is a constant.
We suppose the distance travelled by the \target{} agents depends on how close the herder agents are and model this effect by considering a potential field centred on the $j$-th herder given by
${v}^{(i,j)}={1}/{(\lVert \underline x^{(i)} - \underline y^{(j)} \rVert)}$, exerting on the \target{} agents an action proportional to its gradient \cite{pierson_controlling_2018}.
Specifically, the  dynamics of the $i$-th \target{} agent is influenced by the reaction term
\begin{equation}
\label{eq:Target_repulsion}
V^{(i)}_r(t) = \alpha_r \sum_{j=1}^{N_H} \frac{\partial v^{(i,j)}}{\partial \underline x^{(i)}} = -
\alpha_r \sum_{j=1}^{N_H} \tfrac{\underline x^{(i)} (t) - \underline y^{(j)}(t)}{\|\underline x^{(i)}(t) - \underline y^{(j)}(t)\|^3},
\end{equation}
where $\alpha_r>0$ is a constant. 
Possible modelling uncertainties in the repulsive reaction term \eqref{eq:Target_repulsion} can be seen as being captured by the additional noisy term in \eqref{eq:target_nalepka}.

Notice that according to \eqref{eq:Target_repulsion} every \target{} agent feels the influence of all the herders. Nevertheless, we assume that each herder only chases one target at a time as explained  below.
The position of the $i$-th \target{} agent when it is targeted by the $j$-th herder will be denoted as 
$\underline{\Tilde{x}}^{(i,j)}$ or in polar coordinates as $(\Tilde{\rho}^{(i,j)},\Tilde{\phi}^{(i,j)})$. 
\section{Herder dynamics and control rules}
\label{sec:herder_control}
Our solution to the herding problem consists of two layered strategies; (i)  a local control law to drive the motion of the herder towards the target it selected, and to push it inside the goal region and (ii) a target selection strategy through which herders decide what target to chase.
%\textcolor{violet}{When the herd is contained, the local control law switches autonomously to an oscillatory motion around the goal region. Therefore, the proposed control laws embed both ``collecting'' and ``patrolling'' behaviours of herder agents, independently from herd dynamics.}
When the herd are all gathered, the herders switch back to an idling condition by keeping theirself within the safety region surrounding the goal region.

\subsection{Local control strategy}
For the sake of comparison with the strategy presented in \cite{nalepka_first_2017,Nalepka_2019}, we express in polar coordinates the control law we propose to drive each herder. 
Albeit not resulting in the shortest possible path travelled by the herders, the controller expressed in polar coordinates ensures circumnavigation of the goal region, avoiding targets already contained therein from being scattered around.
Specifically, the control input to the $j$-th herder dynamics \eqref{eq:herder} is defined as $u^{(j)}(t)=u_r^{(j)}(t)\, \hat{r}_j + u_\theta^{(j)}(t)\, \hat{\theta}_j$, where $\hat{r}_j = [\cos \theta^{(j)}, \, \sin \theta^{(j)}]^\top$ and $\hat{\theta}_j = \hat{r}_j^\perp$ are unit vectors, and its components are chosen as
\begin{eqnarray}
\label{eq:control_input_1}
u_r^{(j)}(t) &=& - b_r \dot{r}^{(j)}(t)-\mathcal{R}(\underline{\Tilde{x}}^{(i,j)},t),\\
\label{eq:control_input_2}
u_\theta^{(j)}(t) &=& - b_\theta \dot{\theta}^{(j)}(t)-\mathcal{T}(\underline{\Tilde{x}}^{(i,j)},t) ,
\end{eqnarray}
with $b_r,\,b_\theta>0$, and where the feedback terms $\mathcal{R}(\underline{\Tilde{x}}^{(i,j)},t)$ and $\mathcal{T}(\underline{\Tilde{x}}^{(i,j)},t)$ are elastic forces that drive the herder towards the chased target $i$ and push it towards the containment region $\mathcal{G}$. 
Such forces are chosen as
\begin{equation} 
\begin{split}
	\mathcal{R}(\underline{\Tilde{x}}^{(i,j)},t) = \epsilon_{r} \, \Big[ r^{(j)}(t) & - \xi^{(j)}(t) \, (\Tilde{\rho}^{(i,j)}(t) + \Delta r^{\star}) \\
	& - (1 - \xi^{(j)}(t)) \, ( r^{\star} + \Delta r^{\star}  ) \Big] ,
\end{split}
\end{equation}
\begin{equation}
\label{eq:elastic_torque}
\mathcal{T}(\underline{\Tilde{x}}^{(i,j)},t) = \epsilon_{\theta} \, \left[ \theta^{(j)}(t) - \xi^{(j)}(t) \Tilde{\phi}^{(i,j)}(t) - (1 - \xi^{(j)}(t)) \psi(t) \right] .
\end{equation}
with $\epsilon_{r},\,\epsilon_{\theta}>0$, and where $\xi^{(j)}(t)$ regulates the switching policy between collecting and idling behaviours. That is, $\xi^{(j)}(t) = 1$, if $\Tilde{\rho}^{(i,j)}(t) \geq r^{\star}$, and $\xi^{(j)}(t) = 0$, if $\Tilde{\rho}^{(i,j)}(t) < r^{\star}$, so that the herder is attracted to the position of the $i$-th chased target $\underline{\Tilde{x}}^{(i,j)}$ (plus a radial offset $\Delta r^\star$) when the current target is outside the containment region ($ \xi^{(j)} = 1$) or close to the boundary of the buffer region at the idling position $(r^{\star} + \Delta r^{\star},\, \psi(t))$, in polar coordinates, otherwise ($ \xi^{(j)} = 0$). 
The value of the idling angle $\psi(t)$ depends on the specific choice of the target selection strategy employed, which are discussed next.
Note that the control laws \eqref{eq:control_input_1}-\eqref{eq:control_input_2} are much simpler than those presented in \cite{nalepka_first_2017} as they do not contain any higher order nonlinear term nor are complemented by parameter adaptation rules (see \cite{nalepka_first_2017} for further details). 

\subsection{Target selection strategies}
\label{subsec:TaskDivisionStrategies}

In the case of a single herder chasing multiple agents, the most common strategy in the literature is for it to select the target chased as either the farthest passive agent from the goal region, or the centre of mass of the flocking herd \cite{ vaughan_experiments_2000,strombom_solving_2014, licitra_single_2017}. When two or more herders are involved, the problem is usually solved using a formation control approach, letting the herders surround the herd and then drive them towards the goal region \cite{pierson_controlling_2018, jyh-ming_lien_shepherding_2004}.
Rather than using formation control techniques or solving off-line or on-line optimisation problems as in \cite{jyh-ming_lien_shepherding_2005, Burger2011}, here we present a set of simple, yet effective, target selection strategies that exploit the spatial distribution of the herders allowing them to cooperatively select their targets without requiring any computationally expensive optimisation problem to be solved on-line.

\begin{figure}[!t]
\centering
\subfigure[Global search \label{fig:search_Global}]{%
	\includegraphics[width=0.45\linewidth]{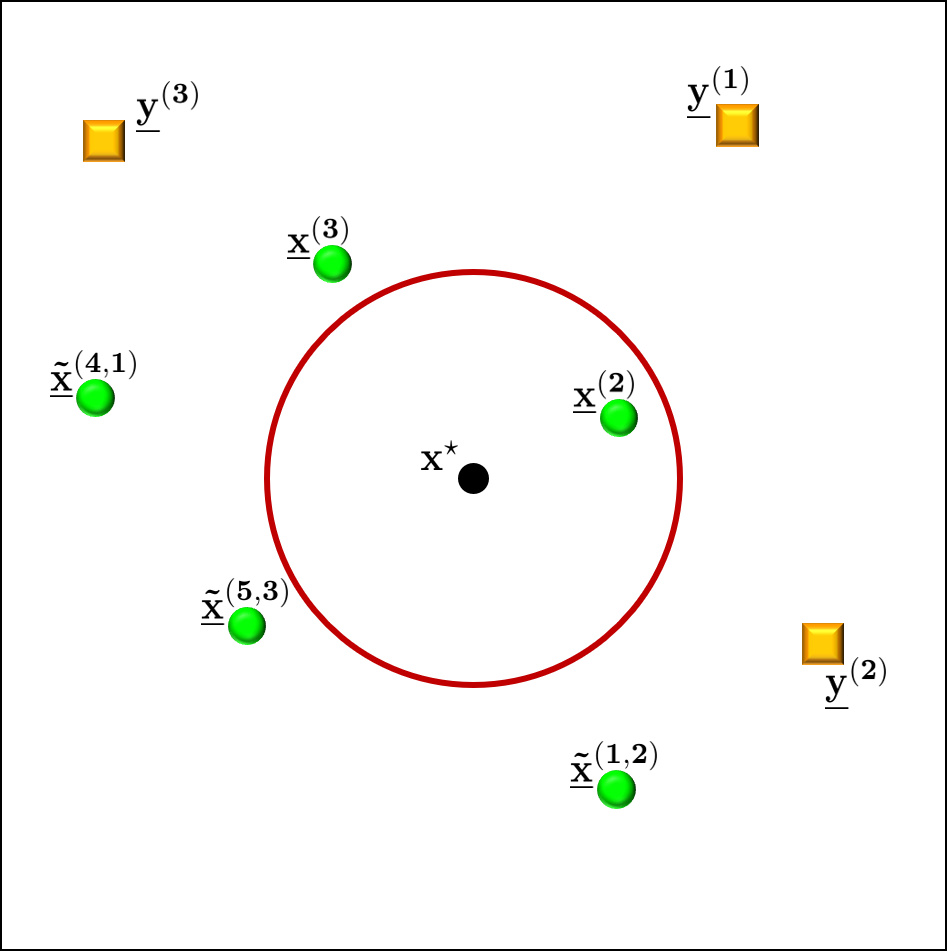}}
\hfill
\subfigure[Static arena partitioning  \label{fig:search_PN}]{%
	\includegraphics[width=0.45\linewidth]{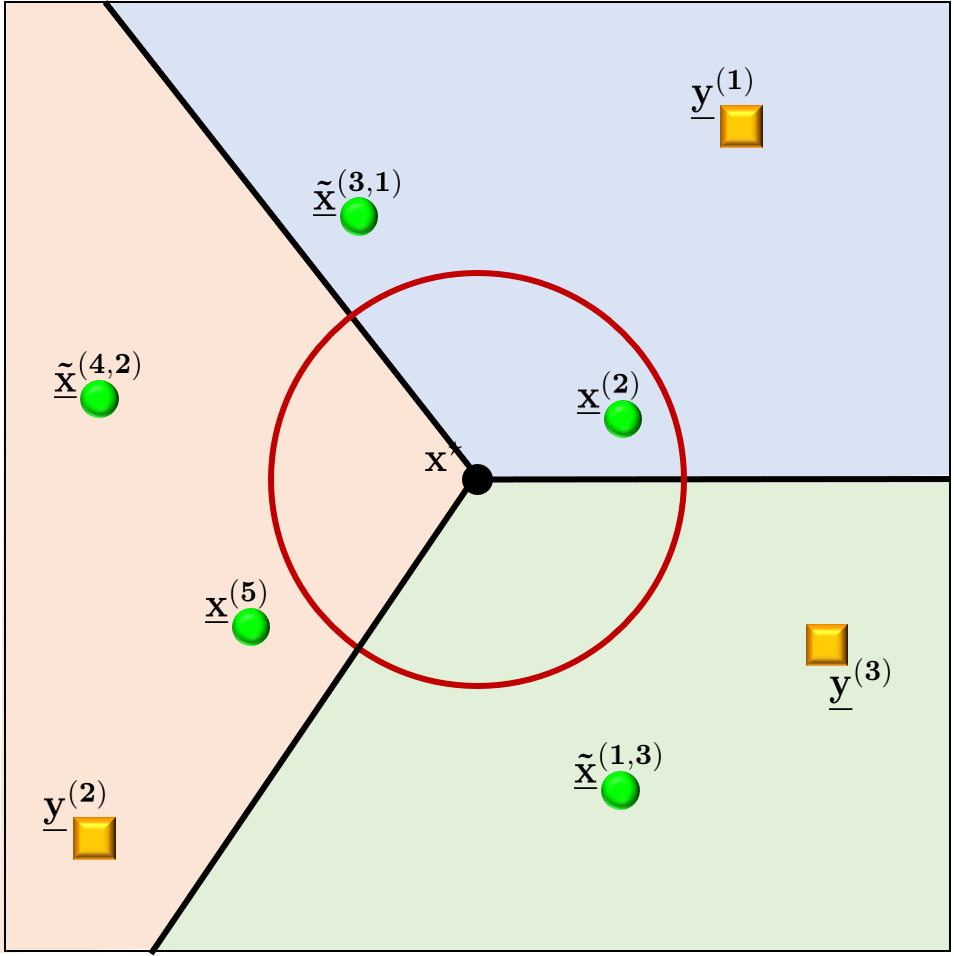}}
\\
\subfigure[Leader-follower  \label{fig:search_DF}]{%
	\includegraphics[width=0.45\linewidth]{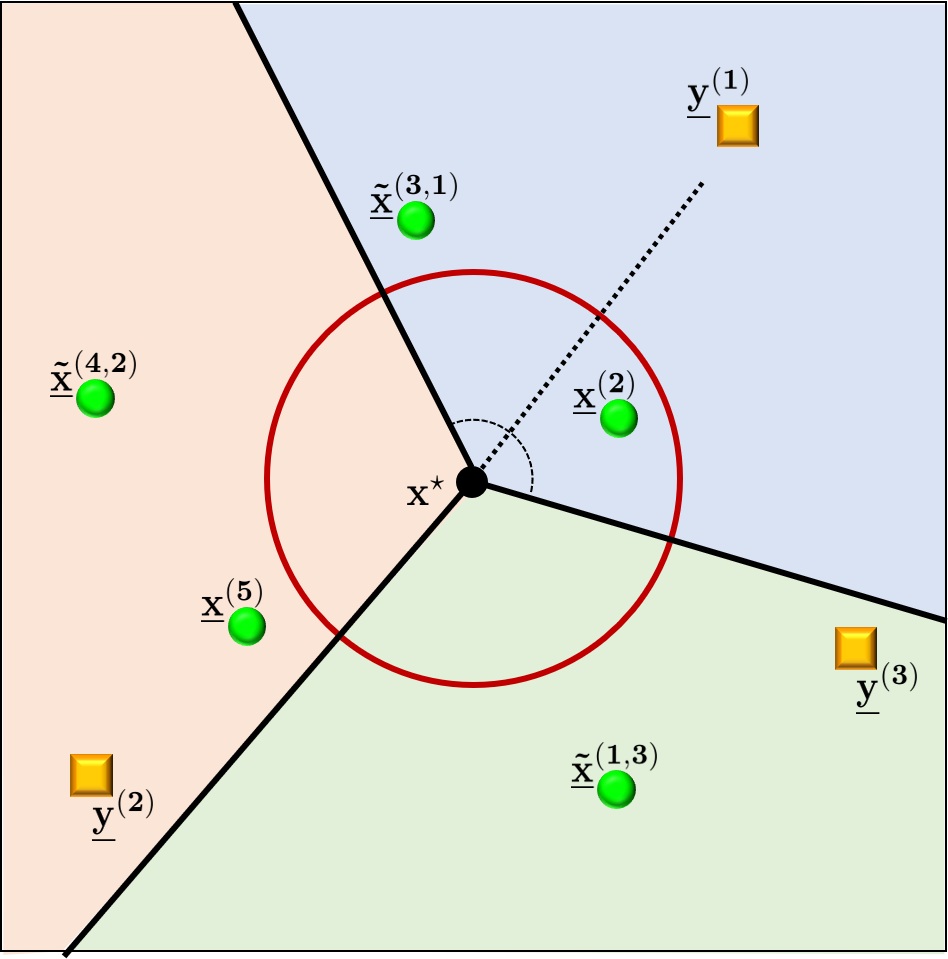}}
\hfill
\subfigure[Peer-to-peer  \label{fig:search_FA}]{%
	\includegraphics[width=0.45\linewidth]{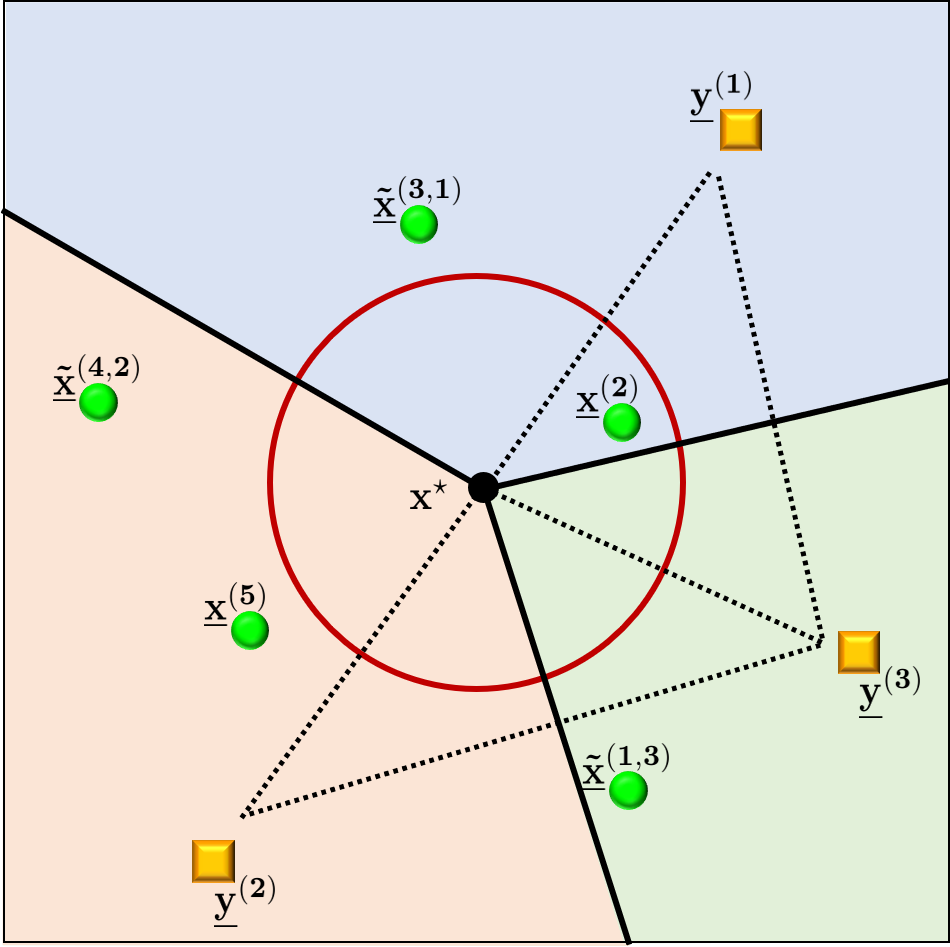}}
\caption{Graphical representation of the target selection strategies. 
	Herders are depicted as yellow squares, \target{} agents as green balls.
	The colours in which the game field is divided correspond to regions assigned to different herders.
	Herder $\underline{y}^{(j)}$ is currently chasing target agent $\underline{\tilde{x}}^{(i,j)}$, while \target{} agent $\underline{x}^{(i)}$ is not chased by any herder.}
\label{fig1} 
\end{figure}

We present four different herding strategies, starting from the simplest case where herders globally look for the target farthest from the goal region.  A graphical illustration of the four strategies is reported in Fig.~\ref{fig1} for $N_H=3$ herders.

\paragraph{Global search strategy (no plane partitioning)} Each herder selects the farthest \target{} agent from the containment region which is not currently targeted by any other herder (Fig.~\ref{fig:search_Global}). 
Being the simplest possible strategy, we will use this strategy as a \emph{benchmark} to compare the performance of the others strategies considered here.

\paragraph{Static arena partitioning} At the beginning of the trial and for all of its duration, the plane is partitioned in $N_H$ circular sectors of width equal to $2 \pi / {N}_H\,\mathrm{rad}$ centred at $\underline{x}^{\star}$. 
Each herder is then assigned one sector to patrol and selects the \target{} agent therein that is farthest from $\mathcal{G}$ (Fig.~\ref{fig:search_PN}).  Note that this is the same herding strategy used in \cite{nalepka_first_2017} for $N_H = 2$ herders.

\paragraph{Dynamic leader-follower (LF) target selection strategy}
At the beginning of the trial, herders are labelled from $1$ to $N_H$ in anticlockwise order starting from a randomly selected herder which is assigned the \emph{leader} role.
The plane is then partitioned dynamically in different regions as follows.
The leader starts by selecting the farthest \target{} agent from $\mathcal{G}$ whose angular position  $\tilde{\phi}^{(i,1)}$ is such that
\begin{equation*}
\tilde{\phi}^{(i,1)} \in \left( \theta^{(1)}(t) - \frac{1}{2}\frac{2\pi}{ N_H},\, \theta^{(1)}(t) + \frac{1}{2}\frac{2\pi}{N_H} \right],
\end{equation*}
where $\theta^{(1)}(t)$ is the angular position of the leader at time $t$. 
Then, all the other \emph{follower} herders ($j=2,\dots,N_H$), in ascending order, select their targets as the passive agent farthest from $\mathcal{G}$ such that
\begin{equation*}
\tilde{\phi}^{(i,j)} \in \bigg(  \theta^{(1)}(t) - \frac{1}{2} \frac{2\pi}{ N_H} + \zeta^{(j)} ,
\theta^{(1)}(t) + \frac{1}{2} \frac{2\pi}{ N_H} + \zeta^{(j)} \bigg],
\end{equation*}
with $\zeta^{(j)} = {2 \pi}(j-1)/{{N}_H}$.
As the leader chases the selected target and moves in the plane, the partition described above changes dynamically so that a different circular sector with \emph{constant} angular width $2\pi/N_H\,\mathrm{rad}$ is assigned to each follower at any time instant. 
In Fig.~\ref{fig:search_DF} the case is depicted for $N_H=3$ in which the sector $(\theta^{(1)}- \frac{\pi}{3}, \theta^{(1)}+ \frac{\pi}{3}]$ is assigned to the leader herder while the rest of the plane is assigned equally to the other two herders.

\paragraph{Dynamic peer-to-peer (P2P) target selection strategy}
At the beginning of the trial herders are labelled from $1$ to $N_H$ as in the previous strategy.
Denoting as $\zeta_j^+(t)$ the angular difference between the positions of herder $j$ and herder $(j+1)\, \mathrm{mod}\, N_H$ at time $t$, and as $\zeta_j^-(t)$ that between herder $j$ and herder $(j+N_H-1)\, \mathrm{mod}\, N_H$ at time $t$,
then herder $j$ selects the farthest \target{} agent from $\mathcal{G}$ whose angular position is such that
\begin{equation*}
\tilde{\phi}^{(i,j)} \in \bigg( \theta^{(j)}(t) - \frac{\zeta_j^-(t)}{2}, \, \theta^{(j)}(t) + \frac{\zeta_j^+(t)}{2} \bigg].
\end{equation*} 
Unlike the previous case, now the width of the circular sector assigned to each herder is also dynamically changing as it depends on the relative angular positions of the herders in the plane.

The idling angle $\psi(t)$ in \eqref{eq:elastic_torque} is set equal to the angular position $\Tilde{\phi}^{(i,j)}$ of the last contained target for the \emph{global search strategy}, otherwise it is set equal to the angular position corresponding to the half of the angular sector assigned at each time to the herder.

A crucial difference between the herding strategies presented above is the nature (local vs global) and amount of information that herders must possess to select their next target.
Specifically, when the {\em global search strategy} is used, every herder needs to know the position $\underline{x}^{(i)}$ of every \target{} agent in the plane, not currently targeted by other herders. In the case of the {\em static arena partitioning} instead a herder needs to know its assigned (constant) circular sector together with the position $\underline{x}^{(i)}$ of every \target{} agent in the sector.

For the dynamic target selection strategies, less information is generally required. Indeed, in the {\em dynamic leader-follower strategy} the herders, knowing $N_H$, can either self-select the sector assigned to themselves (if they act as leader) or self-determine their respective sector by knowing the position of the leader  $\underline{y}^{(1)}(t)$. Similarly in the {\em dynamic peer-to-peer strategy} herders can self-select their sectors by using the angles $\zeta_j^+(t)$ and $\zeta_j^-(t)$.

%\begin{remark}
Note that in the unlikely event of perfect radial alignment of the herder and its target, the herder might push the target away, rather than towards the goal region. Despite its rare occurrence, such an event can be avoided by extending the herder dynamics by extra term \eqref{eq:herder_extra_term} described in \ref{sec:num_simulations}. 
%\end{remark}

\section{Numerical validation} \label{sec:CompleteModel_NumericalValidation}
The herding performance of the proposed control strategies has been evaluated through a set of numerical experiments aimed at (i) assessing their effectiveness in achieving the herding goal; (ii) comparing the use of different target selection strategies; (iii) studying the robustness of each strategy to parameter variations. The implementation and validation of the strategies in a more realistic robotic environment is reported in the next section where ROS simulations are included.

\subsection{Performance Metrics}
We defined the following metrics 
(see  \ref{subsec:metrics} for their definitions) 
to evaluate the performance of different strategies. Specifically, for each of the proposed strategies we computed the (i) gathering time $t_\mathrm{g}$, (ii) the average length $d_\mathrm{g}$ of the path travelled by the herders until all targets are contained, (iii) the average total length $d_\mathrm{tot}$ of the path travelled by herders during all the herding trial, (iv) the mean distance $D_T$ between the herd's centre of mass and the centre of the containment region, and (v) the herd agents' spread $S_{\%}$.

Note that \emph{lower} values of $t_\mathrm{g}$ correspond to \emph{better} herding performance; herders taking a shorter time to gather all the \target{} agents in the goal region. 
Also, \emph{lower} values of $D_T$ and $S_{\%}$ correspond to a \emph{tighter} containment of the \target{} agents in the goal region while  \emph{lower} values of $d_\mathrm{g}$ and $d_\mathrm{tot}$ correspond to a \emph{more efficient} herding capability of the herders during the gathering and containment of the herd.

\subsection{Performance analysis}
\label{sec:performance_analysis}
We carried out 50 simulation trials with $N_T=7$ \target{} agents and either $N_H=2$ or $N_H=3$ herders, starting from random initial conditions. (All simulation parameters and a description of simulation setup adopted here are reported in \ref{sec:num_simulations}.)

\begin{table}[!t]

\centering
\begin{tabular}{|l||c|c|c|c|}
	\hline
	& Global & Static & \makecell{LF} & \makecell{P2P} \\
	\hline
	\multicolumn{5}{|l|}{$N_H=2$} \\
	\hline
	$t_\mathrm{g}$ [a.u.] & 8.52 & 15.19 & 15.31 & 13.34 \\
	$d_\mathrm{g}$ [a.u.] & 139 & 102 & 92 & 143\\
	$d_\mathrm{tot} $ [a.u.]& 841 & 493 & 423 & 418\\
	$D_T $ [a.u.]& 1.26 & 1.44 & 1.46 & 1.29 \\
	$S_\% $ [\%]& 0.15 & 0.18 & 0.21 & 0.21 \\
	%		COC pairs [\%] & 100  & 100 & 2 & 78 \\
	\hline
	\multicolumn{5}{|l|}{$N_H=3$} \\
	\hline
	$t_\mathrm{g}$ [a.u.] & 5.88 & 19.60 & 11.23 & 10.11 \\
	$d_\mathrm{g}$ [a.u.] & 88 & 227 & 84 & 59\\
	$d_\mathrm{tot} $ [a.u.] & 1242 & 814 & 885 & 932\\
	$D_T $ [a.u.]& 0.61 & 1.29 & 0.78 & 0.78\\
	$S_\% $ [\%] & 0.13 & 0.39 & 0.24 & 0.91\\
	\hline
\end{tabular}
\caption{Average performance over 50 successful trials of different herding strategies for $N_T = 7$ \target{} agents.}
\label{tab:herding_Divisions}
\end{table}

The results of our numerical investigation are reported in  Tab.~\ref{tab:herding_Divisions}. 
As expected, when herders search globally for agents to chase, their average gathering and total paths, $d_\mathrm{g}$ and $d_\mathrm{tot}$, are notably longer than when dynamic target selection strategies are used, pointing out that this strategy is going to be the least efficient when implemented.

As regards the aggregation of the herd in terms of $D_T$ and $S_{\%}$, all strategies presented comparable results.
On the other hand, dynamic strategies showed consistently shorter gathering times $t_\mathrm{g}$ and travelled distances $d_\mathrm{g}$ than the static target selection strategies.
In particular, in the case of three herders ($N_H=3$), the peer-to-peer strategy exhibited values of $t_\mathrm{g}$ and $d_\mathrm{g}$ which are $50\%$ and $74\%$ smaller, respectively, than the static partitioning one.
Therefore, we find that in general higher level of cooperation between herders and a more efficient coverage of the plane, as those guaranteed by dynamic strategies, yield an overall better herding performance which is more suitable for realistic implementations in robots or virtual agents that are bound to move at limited speed. 

\subsection{Robustness analysis}

% \begin{figure}[htbp]%[!t]
\begin{figure}[!t]
\begin{center}    
	\subfigure[Gathering time $t_\mathrm{g}$. Lower values correspond to faster herding.
	] {\includegraphics[width=\linewidth]{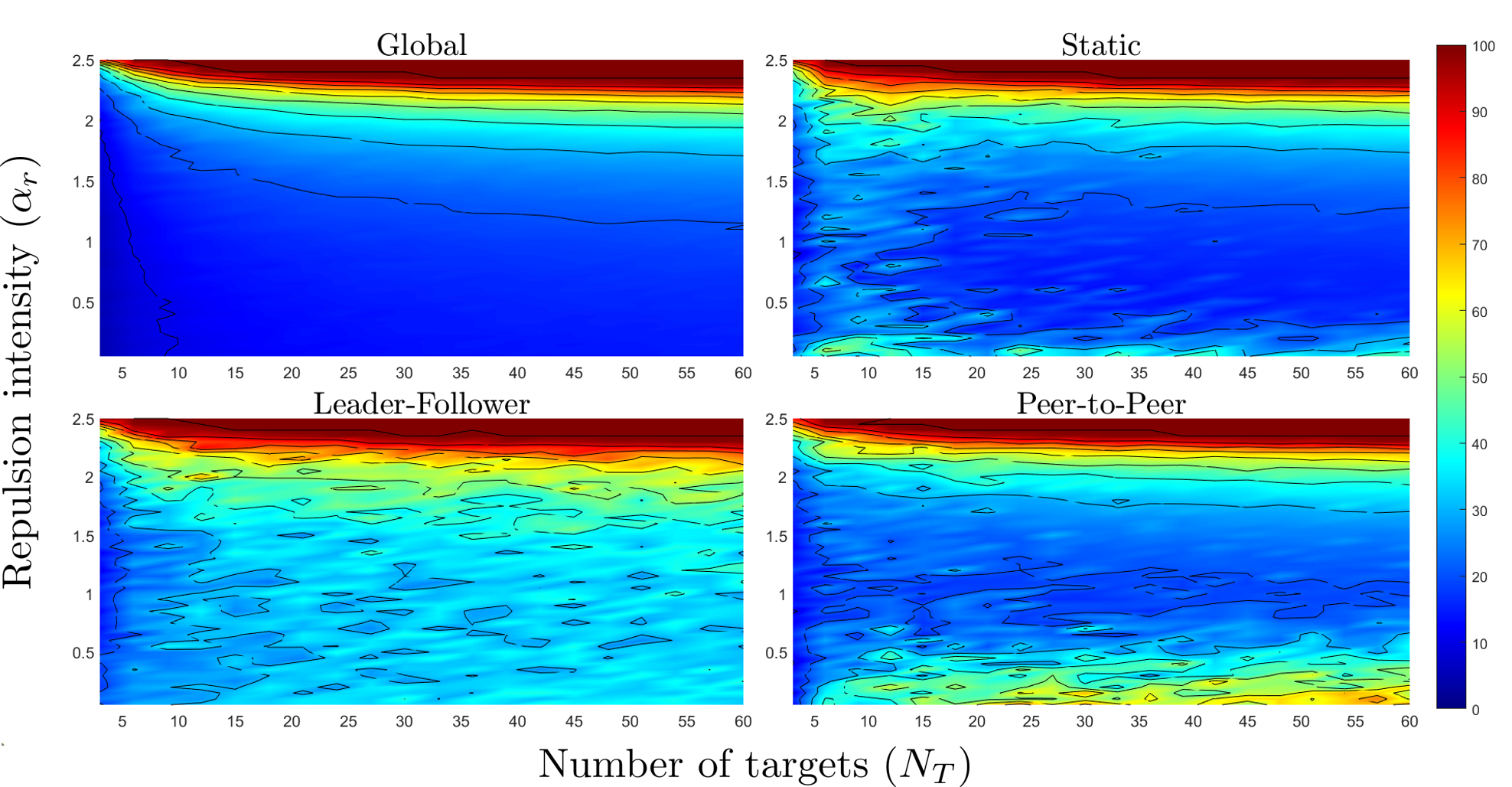}%
		\label{fig:contTimeSurface} } \\
	\subfigure[Total distance travelled $d_\mathrm{tot}$.
	Lower values correspond to more efficient herding.
	]{ \includegraphics[width=\linewidth]{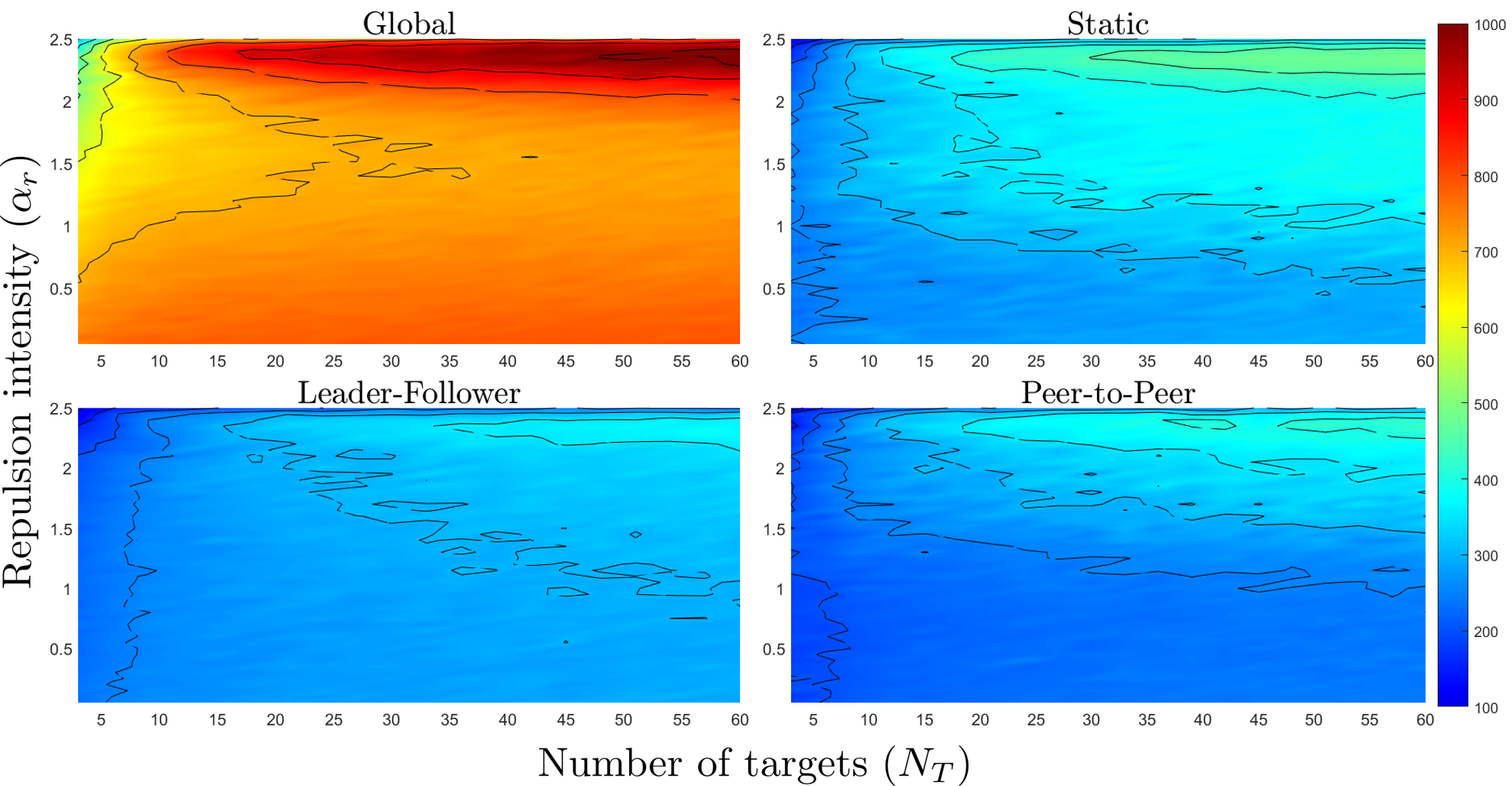}%
		\label{fig:DistanceTravelledSurface} }
	\caption{Robustness analysis of the proposed herding strategies for two herders ($N_H=2$) to variation of herd size $N_T$ and repulsive reaction coefficient $\alpha_r$. 
		$N_T$ was varied between 3 and 60 agents, with increments equal to 3, while $\alpha_r$ between 0.05 and 2.5, with increments equal to 0.05. 
		For each pair ($N_T$, $\alpha_r$) the corresponding metric was averaged over 15 simulation trials starting with random initial positions.
		The coloured plots were obtained by interpolation of the computed values. 
	}
	\label{fig:robustness_analysis}
\end{center}	
\end{figure}

Next, we analysed the robustness of the proposed herding strategies to variations of the herd size and of the magnitude of the repulsive reaction to the herders  exhibited by the \target{} agents (Fig.~\ref{fig:robustness_analysis}).
Specifically, we vary $N_T$ between 3 and 60 and the repulsion parameter $\alpha_r$ in \eqref{eq:Target_repulsion} between 0.05 and 2.5, while keeping $N_H=2$. Strikingly, we find that all strategies succeed in herding up to 60 agents in a large region of parameter values [see the blue areas in Fig.~\ref{fig:contTimeSurface}]. 

The global strategy where herders patrol the entire plane is found as expected to be the least efficient in terms of total distance travelled by the herders (Fig.~\ref{fig:DistanceTravelledSurface});
the dynamic peer-to-peer strategy offering the best compromise and robustness property in terms of containment performance (see Fig.~\ref{fig:contTimeSurface}) and efficiency (see Fig.~\ref{fig:DistanceTravelledSurface}).
To validate these findings we carried out 50 trials where $N_H=3$ herders were required to herd $N_T=60$ \target{} agents, starting from different initial conditions. 
%Over 50 trials with different initial conditions, performance averaged over successful trials are reported in Tab.~\ref{tab:herding_60targets}. 

The resulting performance averaged over the successful trials is reported in Tab.~\ref{tab:herding_60targets}.
Herders adopting the global and peer-to-peer strategies successfully herd all agents in over $50\%$ of the trials. 
Moreover, herders globally searching for the target to chase spent on average slightly less time gathering the targets ($t_g = 12.96$) and achieved and maintained lower herd spread ($S_\%=0.48$), although the path travelled to achieve the goal ($d_\mathrm{tot}$) was significantly higher than when  static or dynamic selection strategies were adopted. 

\begin{table}[!t]
\centering
\begin{tabular}{|l||c|c|c|c|}
	\hline
	& Global & Static & \makecell{LF} & \makecell{P2P} \\
	\hline
	\multicolumn{5}{|l|}{$N_H=3$} \\
	\hline
	\makecell{Successful\\ trials} & 49 & 13 & 8 & 30 \\
	$t_\mathrm{g}$ [a.u.] & 12.96 & 18.22 & 16.06 & 15.94 \\
	$d_\mathrm{g}$ [a.u.]  & 211.04 & 195.47 & 143.53 & 144.92 \\
	$d_\mathrm{tot} $ [a.u.]  & 1226 & 746 & 786.28 & 813 \\
	$D_T $ [a.u.] & 6.5 & 16.92 & 9.7 & 11.99 \\
	$S_\% $ [\%]  & 0.48 & 7.36 & 5.1 & 3.85 \\
	\hline
\end{tabular}
\caption{Average performance over successful trials of different herding strategies for $N_T = 60$ \target{} agents.}
\label{tab:herding_60targets}
\end{table}

\section{Validation in ROS environment}
\label{sec:ROS}
To validate in a more realistic robotic setting the strategies we propose, we complemented the numerical simulation presented in Sec.~\ref{sec:CompleteModel_NumericalValidation} with their ROS implementation\footnote{Code available on \url{https://github.com/diBernardoGroup/HerdingProblem}} as described below.
ROS \cite{ROS} is an advanced software framework for robot software development that provides tools to support the user during all the development cycle, from low-level control and communication to deployment on real robots.
We used the Gazebo software package\footnote{\url{http://wiki.ros.org/gazebo\_ros\_pkgs}} to test the designed control architecture on accurate 3D models of commercial robots to simulate their dynamics and physical interaction with the virtual environment.

\begin{figure}[!t]
\subfigure[Robot agent]{%
	\includegraphics[height=0.195\linewidth]{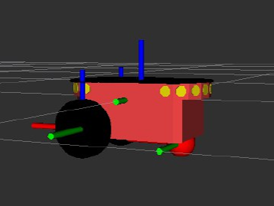}
	\label{fig:ROS_robot}}
\subfigure[Simulated environment]{%
	\includegraphics[height=0.195\linewidth ]{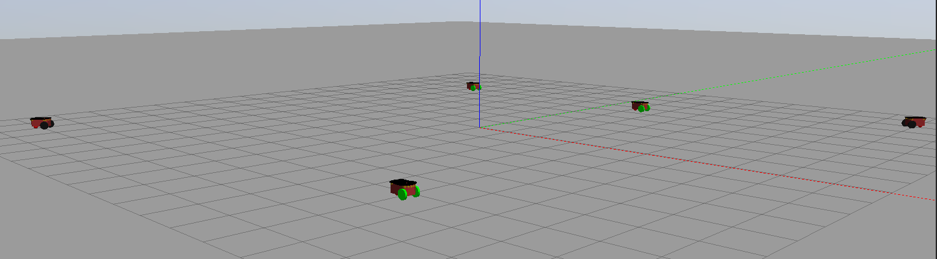}
	\label{fig:ROS_env}}
\caption{Overview of Gazebo-ROS application, with 3D model of the Pioneer 3-DX robot (a) and a landscape view of the simulated environment (b).}
\label{fig:GazeboROS}
\end{figure}

We considered a scenario where $N_T=3$ \target{} agents need to be herded by $N_H=2$ robotic herders. All agents were chosen to be implemented as Pioneer 3-DX \cite{Pioneer3dx}, a commercially available two-wheel two-motor differential drive robot whose detailed model is available in Gazebo (see Fig.~\ref{fig:GazeboROS}). 
%
% \begin{figure}[htbp]
The desired trajectories for the robots are generated by using equations \eqref{eq:target_nalepka} and \eqref{eq:control_input_1}-\eqref{eq:elastic_torque} for the \target{} and herder robots, respectively, which are used as reference signals for the on-board inner control loop to generate the required tangential and angular velocities (see  \ref{app:ros_sims} for further details).

% \begin{figure*}[htbp]%[!t]
\begin{figure*}[!t]
\centering
\subfigure[]{%
	\includegraphics[width=0.3\linewidth]{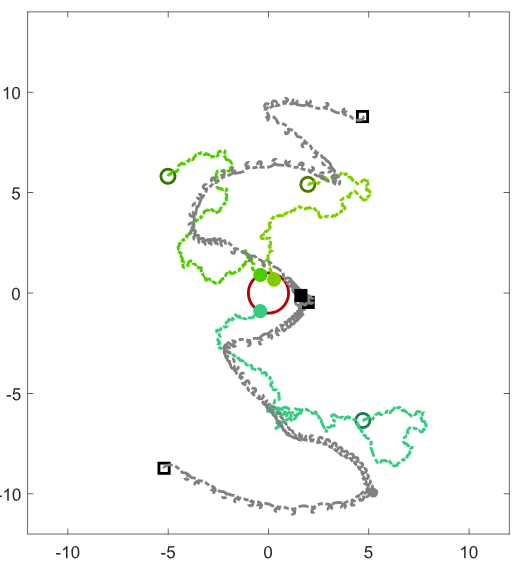}
	\label{fig:GazeboSimulation_static_traj}}
\subfigure[]{%
	\includegraphics[width=0.3\linewidth ]{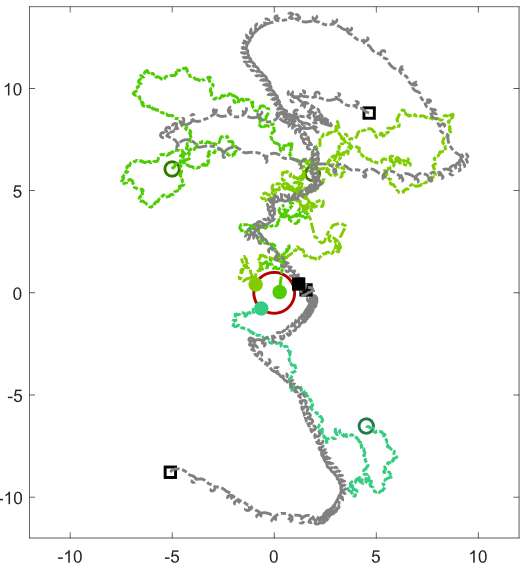}
	\label{fig:GazeboSimulation_leader_traj}}
\subfigure[]{%
	\includegraphics[width=0.3\linewidth]{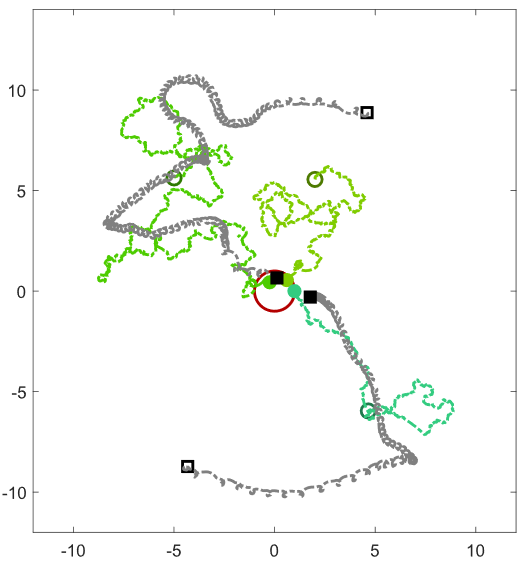}
	\label{fig:GazeboSimulation_p2p_traj}}\\
\subfigure[]{%
	\includegraphics[width=0.3\linewidth]{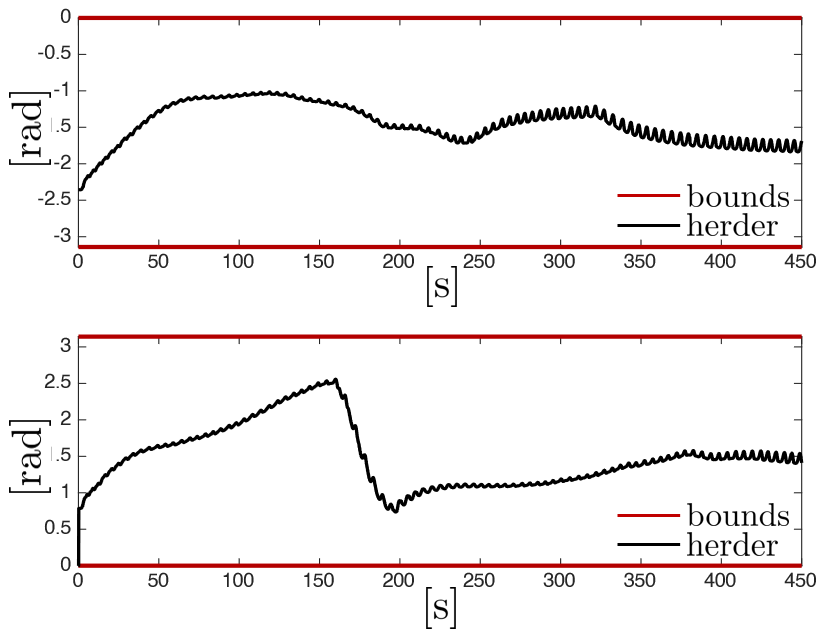}
	\label{fig:GazeboSimulation_static_angle}}
\subfigure[]{%
	\includegraphics[width=0.3\linewidth ]{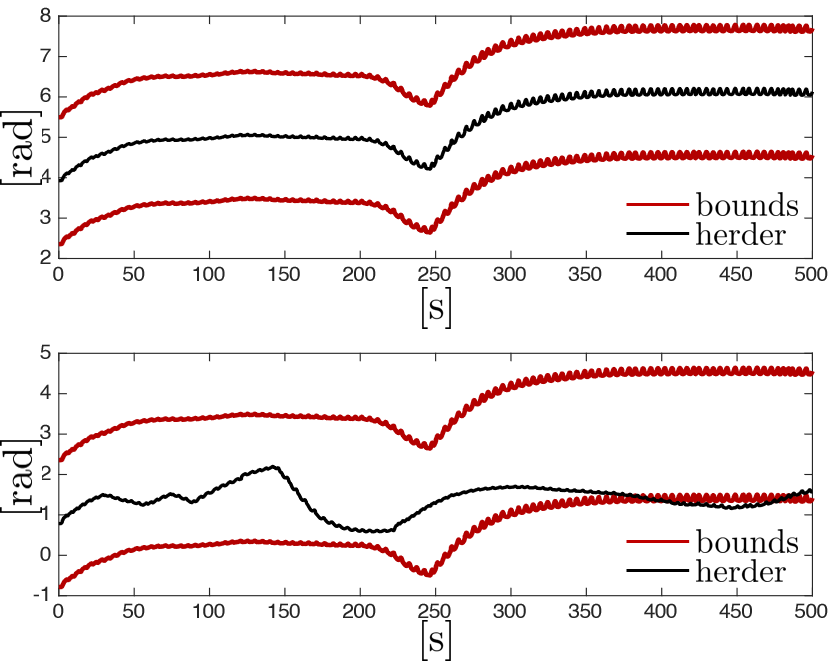}
	\label{fig:GazeboSimulation_leader_angle}}
\subfigure[]{%
	\includegraphics[width=0.3\linewidth]{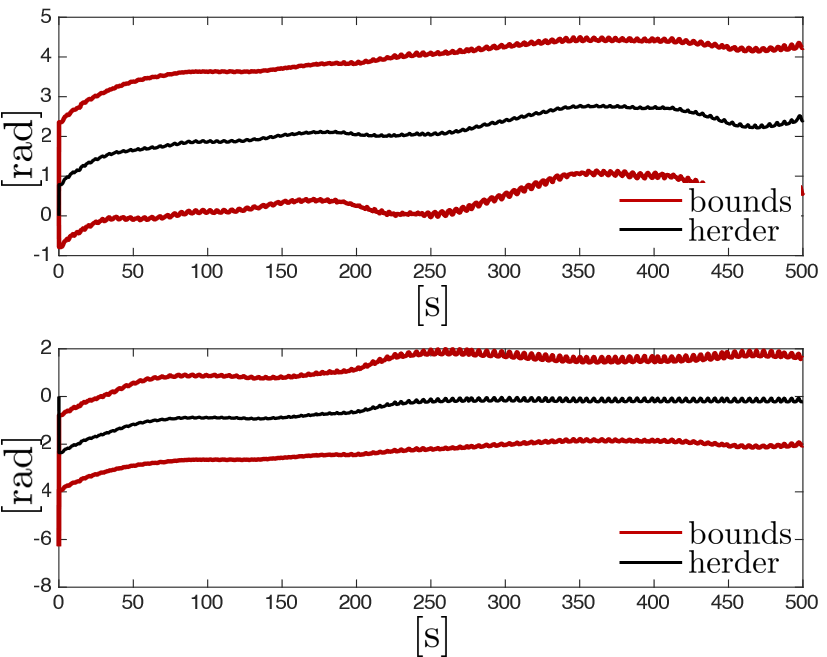}
	\label{fig:GazeboSimulation_p2p_angle}}
\caption{Top panels show the trajectories of \target{} agents (green lines) and herders (black lines) adopting a) static arena partitioning, b) leader-follower and c) peer-to-peer herding strategies simulated in the Gazebo environment. 
	The containment region is depicted as a red circle. 
	Black square marks denote the initial and the final (solid coloured) position of the herders. 
	Green circle marks show the initial and the final (solid coloured) position of the \target{} agents. 
	The value of the herding performance metrics computed for each simulation are also reported on top of the corresponding figures.
	Bottom panels show that all herders are able to collect the herd in less than $500\,\mathrm{s}$ by following the angular bounds (red lines) prescribed by the d) static arena partitioning, e) leader-follower and f) peer-to-peer herding strategies.}
\label{fig:GazeboSimulation}
\end{figure*}

Examples of ROS simulations are reported in Fig.~\ref{fig:GazeboSimulation} where all the target selection strategies that were tested (static arena partitioning, leader-follower, peer-to-peer) were found to be successful with herder robots being able to gather all the  \target{} robots in the containment region. 
Fig.~\ref{fig:GazeboSimulation} also shows that the angular position of the herders remain within the bounds defining the sector of the plane assigned to them for patrolling. 
%The only exception is found in panel Fig.~\ref{fig:GazeboSimulation_leader_angle} where the follower herder temporarily exceeds its bounds.
The only exception is found in panel Fig.~\ref{fig:GazeboSimulation_leader_angle} where the leader-follower strategy is adopted and the follower herder temporarily exceeds the bounds when the leading herder changes its angular position while chasing its target.  This is
essentially due to the subordinate role of the follower herder with
respect to the leader.

\section{Conclusions}
\label{sec:conclusions}
We presented a control strategy to solve the herding problem in the scenario where a group of multiple herders is chasing a group of stochastic \target{} agents. 
Our approach is based on the combination of a set of local rules driving the herders according to the targets' positions and a global herding strategy through which the plane is partitioned among the herders, who then select the target to chase in the sector assigned to them either statically or dynamically. 
Our results show the effectiveness of the proposed strategy both via numerical simulations and by means of a more realistic implementation in ROS on commercially available robotic agents. 
Also, we evaluated the ability of the proposed strategies to cope with an increasing number of \target{} agents and variations of the repulsive force they feel when the herders approach them. 

We wish to emphasise that to date our approach is the only one available in the literature to drive multiple herders to collect and contain a group of multiple agents that do not possess a tendency to flock and whose dynamics is stochastic. 
%All strategies were shown to be robust and effective with the dynamic selection strategies we proposed exhibiting better herding performance and requiring less information to run online.
%Our control approach is therefore an effective and much simpler alternative to other control laws proposed in the literature to solve the herding problem and it also proved to be robust and scalable. 
%A pressing open problem is to derive a formal proof of convergence that given the generality of the approach will be the subject of future work.
A pressing open problem is to derive a formal proof of convergence of the overall control system.
%Interestingly, our numerical evidence suggests that the oscillatory motions of herders observed in experiments with human players \cite{Nalepka_2019} may emerge from the local rules of interaction between herders and target agents and do not need to be explicitly encoded in the mathematical model describing their dynamics. 

\section*{Acknowledgements}
The authors wish to acknowledge support from the Macquarie Cotutelle (Industrial and International Leverage Fund) Award from University of Bristol and International Macquarie University Research Excellence Scholarship Scheme from Macquarie University for supporting Fabrizia Auletta's work. This research was supported, in part, by Australian Research Council Future Fellowship (FT180100447) awarded to Michael Richardson, and in part with the economic support of MIUR (Italian Ministry of University and Research) performing the activities of the  project ARS01\texttt{\char`_}00861 “Integrated collaborative systems for smart factory -  ICOSAF”.
They also wish to thank Dr. Jonathan Cacace from the University of Naples, Italy for his support with ROS.
%

%\section*{References}

\bibliographystyle{elsarticle-num} 
\bibliography{Reference2}

\appendix
\section{Performance Metrics}
\label{subsec:metrics}
Denote with $\mathcal{X}(t):= \left \{ i \, : \, \lVert \underline{x}^{(i)}(t)-\underline{x}^{\star} \rVert \leq r^{\star} \right \}$ the set of \target{} agents which are contained within the goal region $\mathcal{G}$ at time $t$.
Moreover, denote with $[0,T]$ 
the time interval over which the performance metrics are evaluated.
The following metrics are used in the \paper{} to evaluate the proposed herding strategies.\\

{\bf Gathering time} defined as the time instant $t_\mathrm{g}\in[0,\, T]$ such that all the \target{} agents are in the containment region for the first time.\\

{\bf Distance travelled by the herders} which measures the mean in time and among herders of the distance travelled by the herders during the time interval $[0,t]$. It is defined as 
\begin{equation*}
d(t):= \frac{1}{N_H} \sum_{j=1}^{N_H} \frac{1}{t} \left( \int_{0}^{t}  \left\lVert \dot{\underline{y}}^{(j)}(\tau) \right\rVert   d\tau \right) .
\end{equation*}
Therefore, $d_\mathrm{g}:=d(t_\mathrm{g})$, and $d_\mathrm{tot}:=d(T)$. 
A smaller average distance travelled indicates better efficiency of the herders in solving the task.\\

{\bf Herd distance from containment region} which measures the herders ability to keep the herd close to the containment region, with centre $\underline{x}^{\star}$.
It is defined as the mean in time of the Euclidean distance between the centre of mass of the herd and the centre of the containment region, that is
\begin{equation*}
D_T:= \frac{1}{T} \int_{0}^{T}  \left\lVert \left( \frac{1}{N_T} \sum_{i=1}^{N_T} \,  \underline{x}^{(i)}(\tau) \right) - \underline{x}^{\star}(\tau) \right\rVert   d\tau .
\end{equation*}
A smaller average distance indicates better ability of the herders to keep the herd close to the containment region.\\

{\bf Herd spread} measuring how much scattered the herd is in the game field.
Denote as $\mathrm{Pol}(t)$ the convex polygon defined by the convex hull of the points $\underline{x}^{(i)}$ at time $t$, that is, $\mathrm{Pol}(t):= \mathrm{Conv}\left(  \{ \underline{x}^{(i)}(t), \, i=1,\dots,N_T \} \right)$.
Then, the herd spread $S$ is defined as the mean in time of the area of this polygon, that is
\begin{equation*}
S:= \frac{1}{T} \int_{0}^{T} \left( \int_{\mathrm{Pol}(\tau)} d\underline{x} \right) \, d\tau .
\end{equation*}
Lower values corresponds to a more cohesive herd and consequently better herding performance.
The herd spread can also be evaluated with respect to the area of the containment region, $A_\mathrm{cr}=\pi (r^{\star})^2$, as $S_{\%}=S/A_\mathrm{cr}\cdot 100$. 
\section{MATLAB simulations}
\label{sec:num_simulations}
In all simulations we considered the case of ${N}_H = 2$ or ${N}_H = 3$ artificial herders and ${N}_T = 7$ \target{} agents. Moreover, we considered a circular containment region with radius $r^\star = 1$ and a buffer region of width $\Delta r^\star = 1$. 
The numerical integration of the differential equations describing the dynamics of \target{} agents and herders has been realised using Euler-Maruyama method \cite{higham2001algorithmic} in the time interval $[0,T] = \left[0,100\right]\, \mathrm{s}$ with step size $dt=0.006\,\mathrm{s}$.

The values of all parameters used in the simulation were chosen as in \cite{nalepka_herd_2017}. 
Collision detection radius $r_c = 0.0001$, coefficients of  diffusion and repulsive motion $(\alpha_b,\alpha_r)= (0.005,1)$, radial damping and stiffness coefficients $(b_r,\epsilon_r)=(11,98.7)$, angular damping and stiffness coefficients $(b_\theta,\epsilon_\theta)=(11,62.6)$.

The initial positions of the \target{} agents have been set outside the containment region as
$\underline{x}^{(i)}_0 = 2\, r^\star \mathrm{e}^{\jmath \phi^{(i)}_0}$, $\forall i = 1, \dots, {N}_T$,
with $\phi^{(i)}_0$ drawn with uniform distribution in the interval $(-\pi,\pi]$,
while the initial positions of herders have been taken on the circle with radius $4r^\star$ and with angular displacement $(2\pi)/N_H$.
Furthermore, collision avoidance forces between \target{} agents was also considered in the numerical simulations.
Specifically, the model \eqref{eq:target_nalepka} is extended by adding the term $V_c^{(i)}(t) dt$, with 
\begin{equation*}
	V_c^{(i)}(t) = \sum_{ i' \in \mathcal{X}_c^{(i)}(t) }  \frac{\underline x^{(i')}(t) - \underline x^{(i)} (t)}{\| \underline x^{(i') }(t) - \underline x^{(i)} (t) \|^{3}},
\end{equation*}
where $ \mathcal{X}_c^{(i)}(t) :=\{ i' : \lVert \underline{x}^{(i')}(t)-\underline{x}^{(i)}(t) \rVert \leq r_c \}$ is the set of all \target{} agents at time $t$ inside the closed ball centred in $\underline{x}^{(i)}$ with radius $r_c$.

To avoid that perfect alignment between the herder and the chased targeted agent may cause the latter to move away from the goal region, a circumnavigation force $u_{\perp}^{(j)}(t)$ can be added to the dynamics of the herders in \eqref{eq:herder}. This force is orthogonal to the vector $\Delta \underline{x}_{ij} = \underline{x}^{(i)}-\underline{y}^{(j)}$, and its amplitude depends on the angle $\chi_{ij}$ between $\Delta \underline{x}_{ij}$ and $\underline{y}^{(j)}$, such that it is maximum when the two vectors are parallel ($\chi_{ij}=\pi$) and zero when they are anti-parallel ($\chi_{ij}=0$). Specifically, it is defined as: 
\begin{equation}
\label{eq:herder_extra_term}
u_{\perp}^{(j)}(t)=\bar{U} \cdot v(t) \cdot \cos^2 \left( {\frac{\chi_{ij}}{2}} \right) \frac{\Delta \underline{x}_{ij}^{\perp}}{\|\Delta \underline{x}_{ij}\|},
\end{equation}
where $\bar{U}>0$ is the maximum amplitude, and $v \in \{-1,1\}$, whose value depends on which halves of the assigned sector the herder is currently in, to guarantee that the targeted agent is always pushed toward the interior of the sector.
\section{ROS simulations}
\label{app:ros_sims}
%\textcolor{violet}{The Robotic Operating System (ROS) \cite{ROS} provides the framework onto which the robots can collect, elaborate and transmit information. Information, like sensor's measurement or command inputs, are exchanged through a network of ROS nodes. ROS nodes are client libraries, written in C++ or Python by developers, to perform the computations required and to implement the communication paradigm. }
The mobile robots used for both \target{} and herder agents have been designed as Pioneer 3-DX robots driven by the differential drive controller provided in the set of ROS packages (\texttt{gazebo-ros-pkgs}) that allows the integration of Gazebo and ROS.   

The environment and the robots share information through an exchange of messages that occurs publishing and subscribing to one or more of the available topics. 
A ROS node is attached to each herder and \target{} robots. It subscribes to the \texttt{/odom} topic; implements the agent's dynamics; and publishes a personalised \texttt{/cmd\_vel} topic.
The \target{} agents collect odometric information from all the herders in the environment. The herder agents subscribe to the ID of the \target{} agent to-be-chased and collect its position. The published message is a velocity control input w.r.t. the robot's reference system to the differential drive of the robot: a translation $v$ along $x$-axis and a rotation $\omega$ around $z$-axis of the robot. 
The reference trajectory $\underline y^\star(t) =  [r^\star\,\cos \theta^\star, r^\star\,\sin \theta^\star ]^\top$, generated as in Sec.~\ref{sec:TargetModel}-\ref{sec:herder_control}, is followed by each robot by means of the Cartesian regulator 
\begin{eqnarray*} \label{eq:CartesianRegulator}
v & = & -k_1 (\underline y - \underline y^\star) \left[ \cos \Phi, \quad \sin \Phi \right]\\
\omega & =& k_2 (\theta ^\star - \Phi + \pi)
\end{eqnarray*}
where $\Phi(t)$ denotes the robot orientation w.r.t. the global reference system. 
The gains $k_1=0.125$ and $k_2=0.25$ have been tuned by trial-and-error to achieve smooth robot movements.
The initial position of the agents have been set as in \ref{sec:num_simulations}. 

The target selection strategies (Sec.~\ref{subsec:TaskDivisionStrategies}) are processed in an ad-hoc ROS node. It subscribes to the odometry topic; computes the user-chosen strategy (i.e. global, static arena partitioning, leader-follower or peer-to-peer); and publishes a custom message with the ID of the targets to-be-chased on the \texttt{/herder/chased\_target} topic. The custom message is an array of integer numbers, its $j$-th element corresponds to the \target{} agent chased by the $j$-th herder robot. 

The Gazebo-ROS simulations were run on Ubuntu 18.0404 LTS hosted on a VirtualMachine with a 10GB RAM with ROS Melodic distribution and Gazebo 9.13.0.

\end{document}